\documentclass[twocolumn,showpacs,preprintnumbers,amsmath,amssymb]{revtex4}
\usepackage{graphicx}% Include figure files
\usepackage{dcolumn}% Align table columns on decimal point
\usepackage{bm}% bold math

\begin{document}

%\preprint{APS/123-QED}

\title{\boldmath
Impact of Right-handed Strange-beauty Squark on $b \leftrightarrow
s$ Transitions }
\vfill
\author{Wei-Shu Hou}%
%\affiliation{%
%Authors' institution and/or address\\
%This line break forced with \textbackslash\textbackslash
%}%
%
\author{Makiko Nagashima}
% \homepage{http://www.Second.institution.edu/~Charlie.Author}
%\affiliation{${}^a$Institute of Physics, Academia Sinica,
%                 Taipei, Taiwan 115, R.O.C.}
\affiliation{Department of Physics, National Taiwan
University, Taipei, Taiwan 106, R.O.C.
}%
%

%\date{\today}
%
%\vskip -1cm
%
\vfill
\begin{abstract}
As the hint for $CP$ violating new physics in $B\to \phi K_S$ has
weakened, we reconsider the possibility of near maximal mixing
between $\tilde s_R$-$\tilde b_R$ squarks. Such a right-handed
strange-beauty squark $\widetilde{sb}_{1}$ can be realized by
combining supersymmetry with an approximate Abelian flavor
symmetry, and comes with a unique new $CP$ violating phase from
right-handed quark mixing. Naturally heavy strange-beauty squark
and gluino, of order 0.5 to 1 TeV, are easily accommodated by
recent time-dependent $CP$ violation measurements in $B_d\to \phi
K^0$ and $\pi^0 K^0$. Because of near maximal mixing, even with
such heavy masses, the $\widetilde{sb}_{1}$ and $\tilde g$ can
still strongly impact on $B_s$ mass difference and generate $CP$
violation in the mixing, which can still be probed at Tevatron Run
II. But if the scenario is realized, the LHC will provide
definitive information on the new $CP$ phase, and possibly
discover the $\widetilde{sb}_{1}$ squark. Time-dependent $CP$
violation in $B_d \to K^{*0}\gamma$ can be probed at the future
$B$ factory upgrades. Other $b\to s$ decays influenced by large
right-handed dynamics are also discussed.
\end{abstract}
%\keywords{Suggested keywords}%Use showkeys class option if keyword
                              %display desired
\maketitle

%%%%%%%% %%%%%%% %%%%%%% %%%%%%%
\section{Introduction}
%\protect \textbackslash\textbackslash}

The existence of flavor, and the observed patterns associated with
it, are not understood. $CP$ violation (CPV) also seems to be
closely linked to the flavor problem. In the left-handed quark
sector, which can be probed by the weak interaction, information
is summarized in the Cabibbo-Kobayashi-Maskawa (CKM) quark mixing
matrix, which contains a unique CPV phase (usually taken as
$\phi_3/\gamma \equiv \arg\left[V_{ub}^\ast\right]$) if there are
just 3 generations. The CKM matrix elements exhibit a hierarchical
pattern $|V_{us}| \equiv \lambda \simeq 0.22$, $|V_{cb}| \sim
\lambda^2$, $|V_{ub}| \sim \lambda^3 - \lambda^4$, which echo the
hierarchy in quark masses. However, since the weak interaction is
purely left-handed, we have no information on right-handed flavor
physics.

It was observed some time ago that~\cite{Nir,CHPRL01}, if there is
an {\it approximate} Abelian flavor symmetry (AFS) in nature, then
the order of magnitude of the elements of $M_u$ and $M_d$ mass
matrices, in powers of $\lambda$, can be inferred from our current
knowledge of quark masses and mixings. It turns out that the
largest mixing effect, of order 1, would be between the
right-handed $s_R$ and $b_R$ quarks. Such flavor mixing, though
hidden in the Standard Model (SM), would be brought forward if
supersymmetry (SUSY) is also realized. Near maximal mixing between
the $\tilde s_R$ and $\tilde b_R$ squarks would give rise to two
flavor-mixed ``strange-beauty'' squarks $\widetilde{sb}_{1,2}$,
with a single associate new CPV phase defined as
$\sigma$~\cite{CHPRL01,ArhribPRD01}. The strong interaction would
now contain flavor-changing $\widetilde{sb}_{1,2}$-$\{s_R,
b_R\}$-$\tilde g$ couplings, and should affect $b\to s$
transitions.

Sure enough, some ``anomalies" have been uncovered recently in CPV
in $b\to s$ transitions, notably in $B\to \phi K_S$, which has
illustrated the frontier nature of such studies in the past few
years. As the B factories mature, it is exciting that the LHC
would turn on in 2007, making CPV studies involving the $B_s$
system accessible. Furthermore, one can directly search for the
flavor-mixed $\widetilde{sb}_{1,2}$ squarks and probe a broad
range of parameter space.

Time-dependent CPV (TCPV) in $B_d\to J/\psi K_S$ decay (${\cal
S}_{\psi K_S}$) was established in 2001. Because the CKM factor
$V_{cs}^\ast V_{cb}$ for the dominantly tree level $b\to c\bar cs$
transition is expected to be almost real, ${\cal S}_{\psi K_S}$ is
identical to $\sin 2\phi_1/\beta$, the $CP$ phase in $B_d$ mixing
(i.e. $\phi_1/\beta \equiv \arg\left[V_{td}^\ast\right]$) to very
good approximation. The current world average is $\sin
2\phi_1=0.685\pm 0.032$ \cite{HFAG} ($0.73\pm 0.04$ for PDG2005
\cite{PDG}).
TCPV in $B_d\to \phi K_S$ decay, ${\cal S}_{\phi K_S}$, is of
interest because, in SM one expects ${\cal S}_{\phi K_S} \cong
{\cal S}_{\psi K_S}$ to the percent level. This is because the
loop-induced $b\to s\bar ss$ transition that underlies $B_d\to\phi
K_S$ is controlled by the CKM factor $V_{ts}^\ast V_{tb}$, which
again is very close to being real in SM. This makes ${\cal
S}_{\phi K_S}$ an excellent probe of $CP$ violating new physics
(NP).

Interestingly, for the two consecutive years of 2002 and 2003, the
combined result of BaBar and Belle strongly contradicted the SM
prediction of ${\cal S}_{\phi K} \cong \sin 2\phi_1/\beta$. Even
the sign was inconsistent. This so-called {\it sign anomaly}
stimulated many theoretical studies which showed that, to account
for the $\phi K$ sign anomaly, one in general would need large
$s$-$b$ flavor mixing, new CPV phases, and possibly right-handed
dynamics (see, for example,
\cite{rhSUSYKagan,rhSUSYKhalil,rhSUSYKane,rhSUSYHarnik,CHNPRL04}).
This seemed to be an ideal situation for the strange-beauty
squark.
Indeed, our previous work \cite{CHNPRL04} was stimulated by the
2003 result of ${\cal S}_{\phi K}= -0.14\pm 0.33$~\cite{HFAG}. We
showed that a rather light $\widetilde{sb}_1$ and not too heavy
gluino $\tilde g$, such as $[m_{\widetilde{sb}_1},m_{\tilde
g}]\simeq[200,500]$ GeV, together with a large new $CP$ phase
$\sigma\sim 70^\circ$, could give ${\cal S}_{\phi K}<0$. However,
for SUSY scale at TeV, $\widetilde{sb}_1=200$ GeV would require
fine-tuning to ${\cal O}(10^{-2})$ in the squark mass
matrix~\cite{ArhribPRD01}.

Since 2004, however, the experimental discrepancy has weakened
considerably, and we need not adhere to our previous conclusion.
Fine tuning of the $\widetilde{sb}_1$ mass is no longer necessary,
and one could reconsider the model in a more natural setting. The
Belle updated result of ${\cal S}_{\phi K}=0.44\pm 0.27\pm 0.05$
\cite{Bellebsss}, based on 386 million $B\bar B$ events, is in
agreement with BaBar result of $0.50\pm 0.25^{+0.07}_{-0.04}$
based on 227 million $B\bar B$ events \cite{BabarphiK}. The
combined result of ${\cal S}_{\phi K}=0.47\pm 0.19$ \cite{HFAG} is
2.5 $\sigma$ away from the sign anomaly of ${\cal S}_{\phi K}<0$.
But there is still a hint of deviation between ${\cal S}_{\phi
K_S}$ and $\sin 2\phi_1/\beta$, which could be due to
$\widetilde{sb}_1$.

On the other hand, several other modes also show some discrepancy.
The $B\to \pi^0 K_S$ decay involves both $b\to s$ penguin and
$b\to u$ tree contributions. The latter is, however, suppressed by
$V_{us}^\ast V_{ub}/V_{ts}^\ast V_{tb} \lesssim {\cal O}
(V_{us}^2)$, so ${\cal S}_{\pi^0 K_S}\simeq \sin 2\phi_1/\beta$ is
also expected in the SM. From same number of $B\bar B$ events with
${\cal S}_{\phi K_S}$, BaBar finds ${\cal S}_{\pi^0
K_S}=0.35^{+0.30}_{-0.33}\pm 0.04$ \cite{BabarpiK}, and Belle
finds $0.22\pm 0.47\pm 0.08$ \cite{Bellebsss}. The combined result
of $0.31\pm 0.26$ \cite{HFAG} is different from $\sin
2\phi_1/\beta$ with more than 1.4 $\sigma$ significance, and is in
the same direction as ${\cal S}_{\phi K_S}$.

Similarly, ${\cal S}_{\eta^\prime K_S}\simeq \sin 2\phi_1/\beta$
is also predicted in the SM. Yet this is again in some conflict
with the combined result of ${\cal S}_{\eta^\prime K}=0.50\pm
0.09$ \cite{HFAG}. However, the results of BaBar
\cite{BabaretaPRK} and Belle \cite{Bellebsss} are at some odds
with each other, so the error of 0.09 probably should be rescaled
to 0.13. Note that, thanks to the large decay rate, the
measurement of ${\cal S}_{\eta^\prime K_S}$ has better accuracy
compared with ${\cal S}_{\pi^0 K_S}$. However, the large rate of
$B\to \eta^\prime K_S$ is not well understood, and is likely
generated not by new physics.
%besides the NP effects on the TCPV could be diluted very much
%because of its large decay rate \cite{SM4DeltaS}.
In contrast to $B\to \eta^\prime K_S$, the $B\to \pi^0 K_S$ mode
is simpler and more transparent. Hence we focus on the two modes
of $B\to \phi K_S$ and $\pi^0 K_S$ in this work.

The main purpose of this work is to update the picture of
right-handed strange-beauty squarks . We identify new preferred
regions for $m_{\widetilde{sb}_1}$ and $m_{\tilde g}$ from recent
data, and revisit the implications for $B_s$ mixing and the
associated $CP$ violating phase. We believe this update would be
useful for LHC experiments, which would start in 2007.
We follow the observations outlined in Ref.~\cite{HouICHEP04},
which was written after 2004 data revealed drastic softening of
$B\to \phi K_S$ results. With smaller deviations, it is now more
customary to use the difference
\begin{eqnarray}
\Delta{\cal S}_{f}\equiv {\cal S}_{f}-\sin 2\phi_1/\beta,
\end{eqnarray}
which measures the deviation from SM expectations. We put up
useful benchmarks to pin down our model parameters. Since the
error in the $\pi^0 K_S$ data is still large, let us take the
central value of $\phi K_S$ data as a criterion to be more
conservative. We find three scenarios:
\begin{itemize}
\item
\!\!\!{\it Scenario 1}:
$\Delta{\cal S}_{\phi K_S}\simeq -0.22$,
\item
\!\!\!{\it Scenario 2}:
$\Delta{\cal S}_{\pi^0 K_S}\simeq -0.22$,
\item
\!\!\!{\it Scenario 3}:
$\Delta{\cal S}_{\phi K_S,\pi^0 K_S}\simeq  -0.22$,
\end{itemize}
which could provide hint for NP.
In addition, there is a fourth possibility,
\begin{itemize}
\item
\!\!\!{\it Scenario 4}:
$\Delta {\cal S}_{\phi K_S,\pi^0 K_S}\simeq 0$,
\end{itemize}
which, evidently, does not discriminate between NP and the SM.

In this work, we pay attention to the issue of {\it naturalness},
that is, that $\widetilde{sb}_1$ and $\tilde g$ are comparable to
SUSY scale. We shall see that $m_{\widetilde{sb}_1}\gtrsim 500$
GeV and $m_{\tilde g}\gtrsim 700$ GeV can be accounted for by the
recent ${\cal S}_{\phi K_S}$ data. These mass regions are well
within the {\it discovery ranges} at LHC \cite{CMSBrochure}.
Measurements of $B_s$ oscillations and the associated $CP$ phase
will provide further information on $m_{\widetilde{sb}_1}$,
$m_{\tilde g}$ and $\sigma$. In fact, these measurements may
discover NP, even if further B factory results confirm {\it
Scenario 4}.

This paper is organized as follows: Sec.~\ref{sec:CP} concentrates
on the formalism for $CP$ violation observables. In
Sec.~\ref{sec:SM} we give the SM expectations. In
Sec.~\ref{sec:NPmodel} we briefly recapitulate our model of near
maximal $\tilde s_R \leftrightarrow \tilde b_R$ mixing.
Sec.~\ref{sec:result} and \ref{sec:better} gives our NP results.
Discussion and conclusion are given in Sec.~\ref{sec:Epi}. We
refer the hadronic parameters accompanied by the chromo-dipole
effects to Appendix~\ref{app:tildeS}.

%%%%%%%% %%%%%%% %%%%%%% %%%%%%%
\section{\label{sec:CP} \boldmath Time-dependent $CP$ Violation}

In this section we present the formalism for TCPV. For neutral B
meson decays into $CP$ eigenstate $f_{\rm CP}$, the CPV are
studied by means of the time evolution $CP$ asymmetry
\cite{timeCPV},
\begin{eqnarray}
a_{\rm CP}(t)=
%\frac{\Gamma(\bar B_{q}(t)\to f_{\rm CP})-\Gamma(B_{q}(t)\to f_{\rm CP})}
%{\Gamma(\bar B_{q}(t)\to f_{\rm CP})+\Gamma(B_{q}(t)\to f_{\rm CP})}
%\nonumber \\
%&=&
\frac{{\cal A}_{f_{\rm CP}}\cos\Delta m_{B_q}t
+ {\cal S}_{f_{\rm CP}}\sin\Delta m_{B_q} t}
{\cosh\frac{\Delta\Gamma_qt}{2}
+{\cal A}_{\Delta\Gamma}\sinh\frac{\Delta\Gamma_q}{2}}.
\label{eq:timeCPV}
\end{eqnarray}
The coefficients ${\cal A}_{f_{\rm CP}}$, ${\cal S}_{f_{\rm CP}}$
and ${\cal A}_{\Delta\Gamma_q}$ are described in terms of
decay amplitudes ${\cal M}(B_q\to f_{\rm CP})$ and
$\bar {\cal M}(\bar B_q\to f_{\rm CP})$ as
\begin{eqnarray}
& & {\cal A}_{f_{\rm CP}}=
\frac{|\bar\lambda_{\rm CP}|^2-1}{|\bar\lambda_{\rm CP}|^2+1},\;\;\;
{\cal S}_{f_{\rm CP}}=
\frac{2\xi{\rm Im}\bar\lambda_{\rm CP}}
{|\bar\lambda_{\rm CP}|^2+1}, \nonumber \\
& & {\cal A}_{\Delta_\Gamma}=
\frac{2\xi{\rm Re}\bar\lambda_{\rm CP}}
{|\bar\lambda_{\rm CP}|^2+1},
\label{eq:AfSf}
\end{eqnarray}
with $\bar\lambda_{\rm CP}=e^{-i2\Phi_{B_q}}\frac{\bar{\cal
M}}{{\cal M}}$, where $\Phi_{B_q}$ is the $CP$ phase in $B_q$
mixing. The state $f_{\rm CP}$ satisfies ${\cal CP}(f_{\rm
CP})=\xi f_{\rm CP}$. For the $B_d$ system, due to
$\Delta\Gamma_d/\Gamma_d\approx 0$, one finds the simpler form
which is given by
\begin{eqnarray}
a_{\rm CP}(t)= {\cal A}_{f_{\rm CP}}\cos\Delta m_{B_d}t
+ {\cal S}_{f_{\rm CP}}\sin\Delta m_{B_d}t.
\label{eq:timeCPVBd}
\end{eqnarray}
The $CP$ phase $\Phi_{B_d}$ corresponds to $\phi_1/\beta$.

With Wolfenstein parameterization, $\sin 2\phi_1$ can be expressed
in terms of $CP$ violating parameters $\bar\rho\propto
\cos\phi_3/\gamma$ and $\bar\eta\propto \sin\phi_3/\gamma$ as
\cite{Buras98}
\begin{eqnarray}
& & \sin 2\phi_1 = \frac{2\bar\eta (1-\bar\rho)}{(1-\bar\rho^2)^2+\bar\eta^2},
\label{eq:sin2phi1}
\end{eqnarray}
where $\phi_3/\gamma\equiv \arg\left[V_{ub}^\ast\right]$. As we
stated, the value of $\sin2\phi_1$ is basically determined by
${\cal S}_{\psi K_S}$. Although there is a small discrepancy
between unitarity fit of $\sin 2\phi_1$ and direct experimental
result of ${\cal S}_{\psi K_S}$, the approximation $\sin
2\phi_1\approx {\cal S}_{\psi K_S}$ remains reasonable. Therefore,
in order to evade uncertainties brought from $J/\psi K_S$ decay
amplitude, we use Eq.~(\ref{eq:sin2phi1}) for calculation of $\sin
2\phi_1$. In what follows, we take
$\left|{V_{ub}}/{V_{cb}}\right|=0.09$ and $\lambda\equiv
V_{us}=0.22$. We then find $\sin 2\phi_1\simeq 0.73$ for
$\phi_3=60^\circ$, and $\sin 2\phi_1\simeq 0.69$ for
$\phi_3=47^\circ$.

Impact of $CP$ violating NP on TCPV in $b\to s$ transitions can be
understood as follows. Let us take $B\to \phi K$ arising from
purely $b\to s\bar ss$ as an example. We write its amplitude as
${\overline{\cal M}} = a\,e^{i\delta_a} + b\,e^{i\delta_b} \,
e^{i\Phi_{\rm new}}$. The first term denotes the SM contribution
which contains the strong phase $\delta_a$ but approximately no
weak phase because of ${\rm Im}[V_{ts}^\ast V_{tb}]\approx 0$. The
second term, instead, corresponds to the NP contribution with
strong phase $\delta_b$, accompanied by a new $CP$ phase
$\Phi_{\rm new}$. One finds the simple expression for $\Delta
{\cal S}_{\phi K_S}$ up to ${\cal O} (b/a)$
\cite{SfGronau,rhSUSYKhalil}
\begin{eqnarray}
\Delta{\cal S}_{\phi K_S} =
\frac{-\frac{2b}{a}\cos\delta\cos2\phi_1\sin\Phi_{\rm new}}
{1+\frac{2b}{a}\cos\delta\cos\Phi_{\rm new}},
%\Delta{\cal S}_{\phi K_S} \propto \frac{2b}{a}\sin(2\phi_1-\Phi)
%+ \frac{b^2}{a^2}\sin 2(\phi_1-\Phi).
\label{eq:approxDeltaS}
\end{eqnarray}
where $\delta=\delta_b-\delta_a$. $\Delta{\cal S}_{\phi K_S}$
basically follows $\sin\Phi_{\rm new}$ around zero. We see from
Eq.~(\ref{eq:approxDeltaS}) that, unless $\Phi_{\rm new}\neq 0$,
$\Delta{\cal S}_{\phi K_S}$ vanishes for any additional
contributions that carry only strong phases. Therefore, either SM
or NP without CPV mechanism, {\it i.e.} $b=0$ or $\sin\Phi_{\rm
new}=0$, would give $\Delta{\cal S}_{\phi K_S}=0$. However, we
note that Eq.~(\ref{eq:approxDeltaS}) could be diluted by the
relative strong phase $\delta$. For $\delta=90^\circ$ as a typical
case, any NP effect is washed away in $\Delta{\cal S}_{f}$ (one
would then in general get large $\Delta{\cal A}_{f}$, which is not
the case).

To calculate decay amplitudes we need to evaluate hadronic matrix
elements. It is known that the naive factorization (NF) framework
\cite{NF}, which involves small strong phase, has difficulties in
explaining current experimental measurements on decay rates and on
direct CP violation. Recent developments of factorization
frameworks, QCDF \cite{QCDF} and PQCD (see, for example,
\cite{piKPQCD,pipiPQCD}), have shown the importance of
annihilation processes which can generate sizable strong phases
(but not at $90^\circ$ level). Such annihilation effects could
manifest differently in SM and NP, and in different modes. Our
interest is the genuine effects on $\Delta{\cal S}_{\phi K_S}$
from NP. Because of the small strong phases, the analysis within
NF framework would be rather transparent. In the following we use
NF in our calculation and assume absence of final state
interactions. We will illustrate the possible dilution from
presence of additional strong phases by introducing a heuristic
term.

%%%%%%%% %%%%%%% %%%%%%% %%%%%%%
\section{\label{sec:SM}Standard Model Expectations}

This section is devoted to SM calculations. Besides following
Ref.~\cite{NF}, we will take into account the {\it so-called}
chromo-dipole effects from $b\to sg$ \cite{ChromoDipole}. For
hadronic decays from $b\to s$ transition, the effective
Hamiltonian is given by
\begin{eqnarray}
& & \!\!H_{\rm eff}
= \frac{G_F}{\sqrt 2}
\biggl\{V_{us}^\ast V_{ub}C_{1,2} O^u_{1,2}
%+ C_{2} O^u_{2} + C^\prime_{1} O^{\prime u}_{1}
%+ C^\prime_{2} O^{\prime u}_{2}
%\nonumber \\
%& &
+V_{cs}^\ast V_{cb}C_{1,2} O^c_{1,2}
%+ C_{2} O^c_{2} + C^\prime_{1} O^{\prime c}_{1}
%+ C^\prime_{2} O^{\prime c}_{2}
\nonumber \\
& & -V_{ts}^\ast V_{tb}\biggl(\sum_{i=3}^{10} (C_{i}O_i+
C_{i}^{\prime}O_i^\prime) + C_8^G O_8^G+ C_{8}^{\prime
G}O_{8}^{\prime G} \biggr)\biggr\}.
\nonumber \\
\end{eqnarray}
One has \cite{NF} current-current, strong penguin and
electroweak penguin operators, $O_{1,2}^{u,c}$,
$O_{3}$-$O_6$ and $O_{7}$-$O_{10}$.
In addition, the chromo-dipole operator $O_{8}^{G}$ is defined
as
\begin{eqnarray}
O_{8}^G=\frac{\alpha_s}{2\pi}(T^A_{ij}T^A_{kl})\frac{m_b}{q^2}
i\left[\bar s_i\sigma_{\mu\nu}(1+\gamma_5)q^\nu b_j\right]
\left[\bar q^\prime_k \gamma^\mu q^\prime_l\right],
\end{eqnarray}
where $T^A_{ij}$ ($A=1,\cdots 8$) is a generator of SU(3) with color indices $ij$, and
$q$ denotes the momentum carried by the virtual gluon.
The operators $O_i^\prime$ arise from
right-handed dynamics, which are represented by
exchanging  $L\leftrightarrow R$ everywhere.
Since the weak interaction probes only left-handed dynamics, the
short-distance coefficient $C_i^\prime$ in SM is suppressed by a
factor of $m_s/m_b$. In what follows, we will neglect the SM
contributions in the primed coefficients.

%%%%%%%% %%%%%%% %%%%%%% %%%%%%%
%
% table I
%
\begin{table}[t]
\caption{\label{tab:Wilson} Numerical values of $a_i$ for $b\to s$
$(\bar b\to \bar s)$ transition which is obtained by a combination
of the effective Wilson coefficients. In calculating the effective
Wilson coefficients, we follow Ref.~\cite{ChenPRD99} and take
$\phi_3=60^\circ$ and $q^2=m_b^2/3$, where $q^2$ represents the
momentum squared carried by the virtual gluon. Note that
$a_3$-$a_{10}$ are in unit of $10^{-4}$. }
\begin{ruledtabular}
\begin{tabular}{ccc}
$a_i$ & $b\to s$ & $\bar b\to\bar s$ \\ \hline
$a_1$ & $1.046$ & $1.046$ \\
$a_2$ & $0.024$ & $0.024$ \\
$a_3$ & $72.24$ & $72.24$ \\
$a_4$ & $-401.8+32.29\;i$ & $-395.8+33.66\;i$\\
$a_5$ & $-27.57$ & $-27.57$ \\
$a_6$ & $-454.7+i32.29\;i$ & $-448.7+33.66\;i$\\
$a_7$ & $-1.447-0.043\;i$ & $-1.342-0.020\;i$ \\
$a_8$ & $3.076-0.014\;i$ & $3.111-0.007\;i$ \\
$a_9$ & $-93.02-0.043\;i$ & $-92.92-0.020\;i$ \\
$a_{10}$ & $0.129-0.014\;i$ & $0.164-0.007\;i$\\
\end{tabular}
\end{ruledtabular}
\end{table}

%%%%%%% %%%%%%% %%%%%%% %%%%%%%

Let us start from $B\to \phi K_S$. The decay $\bar B^0\to \phi
\bar K^0$ does not occur at tree level. This decay had been
studied within QCDF \cite{phiKQCDF,QCDF} and PQCD framework
\cite{phiKPQCD}, giving decay rate in agreement with the
experimental result of ${\cal B}\simeq (7-9)\times 10^{-6}$
\cite{HFAG}.

With NF,
the amplitude for $\bar B^0\to \phi \bar K^0$ is given by,
\begin{eqnarray}
\frac{{\cal M}(\phi \bar K^0)}{\epsilon^\ast\cdot P_B}
 &=& -\kappa_{\phi K}\;\Biggl\{a_3+a_4+a_5 \nonumber \\
 & & \hspace{-1cm} -\frac{1}{2}\left(a_7+a_9+a_{10}\right)
  +C_{8}^{G}\frac{\alpha_s}{4\pi}\frac{m_b^2}{q^2}\tilde{S}_{\phi K}\Biggr\}\;,
 \label{eq:ampphik}
\end{eqnarray}
where $\kappa_{\phi K} = \sqrt{2}G_F f_\phi m_\phi F_{BK} V_{ts}^*
V_{tb}$, and $\epsilon^\ast\cdot P_B$ is the scalar product of the
$\phi$ meson polarization vector with $B$ meson momentum. The
definition of the parameters $a_i$ can be found in Ref.~\cite{NF}.
Table~\ref{tab:Wilson} enumerates the numerical values of $a_i$
which we use for calculation. We take $C_{8}^G(m_b)= -0.15$ that
would be derived from $b\to s g$. The hadronic parameter
$\widetilde{S}_{\phi K}/q^2$ is taken for $q^2=m_b^2/3$, and
$\widetilde S_{\phi K}\simeq -1.32$ from the evaluation in NF,
which can be found in the Appendix.

Taking $f_\phi=237$ MeV and $F_{BK}(0)=0.35$, we find ${\cal
B}_{\rm SM}(\phi K^0)\simeq 2.4\times 10^{-6}$. Without the
chromo-dipole effects, i.e. $\widetilde S_{\phi K}\to 0$, one has
${\cal B}_{\rm SM}(\phi K^0)\simeq 5.3\times 10^{-6}$. Due to
$\widetilde S_{\phi K}\sim {\cal O} (1)$, the chromo-dipole
contribution is substantial, and gives large reduction for the
rate. But even for $\widetilde S_{\phi K}\to 0$, the decay rate
obtained by NF calculation is far below the experimental data.

For the the decay rate, NF seems deficient.
However, the SM result for ${\cal S}_{\phi K_S}$ is independent
of the factorization framework.
As we noted, any additional effects do not change
Eq.~(\ref{eq:approxDeltaS}) unless an extra CPV phase enters.
Taking account of ${\rm Im}[V_{ts}^\ast/V_{ts}]$, we find
$\Delta S_{\phi K_S}\simeq 0.02$, for $\phi_3/\gamma=60^\circ$,
which is in good agreement with Refs.~\cite{BenekePLB05,FSIPRD05}.

We turn to $B\to \pi^0K^0$. For the decay rate, the QCDF result
\cite{QCDF} and the PQCD result \cite{piKPQCD} are comparable to
the current data ${\cal B}\sim (11-13)\times 10^{-6}$ \cite{PDG}.
In the NF framework, the $\bar B^0\to\pi^0 \bar K^0$ amplitude is
\begin{eqnarray}
{\cal M}(\pi^0 \bar K^0) &=& \kappa_{\pi K}^{(BK)}\Biggl[
\frac{V_{us}^\ast V_{ub}}{V_{ts}^\ast V_{tb}}
a_2 -\frac{2}{3}(a_9-a_7) \Biggr] \nonumber \\
& + &\kappa_{\pi K}^{(B\pi)} \Biggl[ a_4 + \frac{m_K^2(2 a_6
-a_8)}{(m_s+m_d)(m_b-m_d)}
\nonumber \\
& & \hspace{7mm}
 -\frac{1}{2}a_{10} + C_{8}^G \frac{\alpha_s}{4\pi}\frac{m_b^2}{q^2}
\widetilde S_{\pi K} \Biggr], \label{eq:amppik}
\end{eqnarray}
where $\kappa_{\pi K}^{(BK)} = -i\frac{G_F}{2}
f_{\pi}(m_B^2-m_K^2) F_{BK}V_{ts}^\ast V_{tb}$ for the $B\to K$
transition, while $\kappa_{\pi K}^{(B\pi)} = -i\frac{G_F}{2} f_K
(m_B^2-m_\pi^2) F_{B\pi}V_{ts}^\ast V_{tb}$ for the $B\to \pi$
transition. The chromo-dipole contribution appears only in the
latter, and we shall use $\widetilde S_{\pi K}\simeq -1.55$.
Taking $f_\pi=132$ MeV and $m_s=110$ MeV, Eq.~(\ref{eq:amppik})
leads to ${\cal B}_{\rm SM}(\pi^0 K)\simeq 2.8\times 10^{-6}$.
Just as $\phi K$, $\widetilde S_{\pi K}\sim {\cal O} (1)$ reduces
the rate substantially. But even ${\cal B}_{\rm SM}(\pi^0 K)\simeq
4.0\times 10^{-6}$ for $\widetilde S_{\pi K}\to 0$ remains
problematic.

For $\Delta{S}_{\pi^0 K_S}$, we are not hampered by using NF.
Eq.~(\ref{eq:amppik}) is more complicated than
Eq.~(\ref{eq:ampphik}). Furthermore, $\pi^0 K$ would be smeared
with the tree contributions carrying ${\rm Im}[V_{ub}]$ although
$V_{ub}$ is highly suppressed. However,
Eq.~(\ref{eq:approxDeltaS}) persists. The amplitude in
Eq.~(\ref{eq:amppik}) gives $\Delta S_{\pi^0 K_S}\simeq 0.03$, for
$\phi_3/\gamma=60^\circ$, again in agreement with
Refs.~\cite{BenekePLB05,FSIPRD05}.

In similar way, we evaluate TCPV observables in $B \to \eta^\prime
K_S$, $\omega K_S$, $\rho^0 K_S$ modes. The amplitudes of these
modes are rather complicated. But the trend of our results are
consistent with the results in Refs.~\cite{BenekePLB05,FSIPRD05}.
For $\phi_3/\gamma=60^\circ$, we find $\Delta{\cal S}_{\eta^\prime
K_S}\simeq 0.02$, $\Delta{\cal S}_{\omega K_S}\simeq 0.08$, and
$\Delta{\cal S}_{\rho^0 K_S}\simeq -0.01$, respectively. We also
calculate TCPV observables in $\pi^0K^\ast$ and $\eta^\prime
K^\ast$. We assume these final states to be the CP even state, via
$K^\ast \to \pi^0 K_S$. As anticipated, we find ${\cal S}_{\pi^0
K^\ast}\simeq -0.76$ and ${\cal S}_{\eta^\prime K^\ast}\simeq
-0.75$, respectively, with minus sign coming from $\xi$.
The relevant chromodipole hadronic parameters $\widetilde S_{PP}$,
$\widetilde S_{PV}$ and $\widetilde S_{VV}$ can be found in
Appendix A.

%%%%%%%% %%%%%%% %%%%%%% %%%%%%%
\section{\label{sec:NPmodel} \boldmath
Strange-beauty $\widetilde{sb}_R$ squarks}

In this section, we briefly summarize our model without going into
details.
The left-handed flavor mixing is well understood in terms of the
CKM matrix elements. The pattern of flavor mixing in left- and
right-handed dynamics could be different. But within the SM we are
unable to probe right-handed quark mixing.

Our model is one of the possibilities to address $CP$ violation in
the right-handed sector. If there is an underlying approximate
Abelian flavor symmetry (AFS), the right-handed $s$ and $b$ quarks
can have near maximal mixing \cite{Nir,CHPRL01}. The effective AFS
implies down-type quark mass matrix to be of the form
\cite{ArhribPRD01},
\begin{eqnarray}
\frac{M_d^{sb}}{m_b}\sim \left[
\begin{array}{cc}
%\lambda^2 & \lambda^2 \\
\frac{m_s}{m_b} & \frac{m_s}{m_b} \\
1 & 1
\end{array}
\right],
\end{eqnarray}
where we quote 2-3 sector only. The ratio $m_s/m_b$ is
approximately ${\cal O} (\lambda^2)$. The near maximal
right-handed mixing may be the largest off-diagonal element.
However, its effect is hidden from our view in the SM for absence
of right-handed dynamics. The combination of AFS and SUSY brings
forth right-handed dynamics involving squarks, as well as
realizing a near maximal $\tilde s_R$-$\tilde b_R$ squark mixing
\cite{CHPRL01,ArhribPRD01}.

As pointed out in \cite{CHPRL01,ArhribPRD01},
applying four texture zeros
is needed to be safe from kaon low energy constraints.
Decoupling $d$ flavor is implied by lack of NP indication in $B_d$ mixing.
From this backdrop, we focus on 2-3 generation subsystem.
With decoupling of $d$ flavor, the down squark mass matrix
is reduced from $6\times 6$ to $4\times 4$, which is split up into
\begin{eqnarray}
(\widetilde M_d^2)^{(sb)} =
\Biggl[
\begin{array}{cc}
(\widetilde M_d^2)^{(sb)}_{LL} &
(\widetilde M_d^2)^{(sb)}_{LR} \\
(\widetilde M_d^2)^{(sb)}_{RL} &
(\widetilde M_d^2)^{(sb)}_{RR}
\end{array}
\Biggr],
\end{eqnarray}
where each submatrix would in principle be complex, and the
Hermitian nature implies $(\widetilde
M_d^2)^{(sb)\dagger}_{LR}=(\widetilde M_d^2)^{(sb)}_{RL}$. With
squark mass scale $\tilde m$, one finds $(\widetilde
M_d^2)^{(sb)}_{LL}\sim \tilde m^2 V^{(23)}_{\rm CKM}$ and
$(\widetilde M_d^2)^{(sb)}_{LR}\sim \tilde m M^{sb}_{d}$, while
near democratic structure in $(\widetilde M_d^2)^{(sb)}_{RR}\sim
\tilde m^2$ \cite{Chua01,ArhribPRD01}.

We now parameterize $(\widetilde M_d^2)^{sb}_{RR}$ as
\begin{eqnarray}
(\widetilde M_d^2)^{(sb)}_{RR} =
\Biggl[
\begin{array}{cc}
\widetilde m^2_{22}& \widetilde m^2_{23}e^{-i\sigma} \\
\widetilde m^2_{23}e^{i\sigma} & \widetilde m^2_{33}
\end{array}
\Biggr],
\label{eq:massrr}
\end{eqnarray}
with a unique new $CP$ phase $\sigma$, which cannot be rotated
away since phase freedom is already used in quark sector.
Transforming
\begin{eqnarray}
\left(\begin{array}{c}
\widetilde{sb}_1\\
\widetilde{sb}_2
\end{array}
\right) &=& R\left(\begin{array}{c} \tilde s_R \\ \tilde b_R
\end{array}
\right)\nonumber \\
&=& \Biggl[
\begin{array}{cc}
\cos\theta & -\sin\theta e^{-i\sigma} \\
\sin\theta & \cos\theta e^{-i\sigma} \end{array}
\Biggr]\left(
\begin{array}{c}
\tilde s_R \\ \tilde b_R
\end{array}
\right),
 \label{eq:masseigen}
\end{eqnarray}
one obtains the mass eigenstates $\widetilde{sb}_{1,2}$, with the
corresponding eigenvalues $\tilde m^2_{1,2}= (\tilde
m^2_{22}+\tilde m^2_{33} \mp \sqrt{(\tilde m^2_{22}-\tilde
m^2_{33})^2 +4\tilde m^4_{23}})/2$. In general, with $\tilde
m_{22,33}^2\simeq \tilde m_{23}^2$, strange-beauty squark
$\widetilde{sb}_1$ could be made lighter as a consequence of level
splitting.

We assume that SUSY is at TeV scale, and the AFS scale is not too
far above the SUSY scale. To make $\widetilde{sb}_1$ as light as
200 GeV for TeV scale SUSY, one needs fine-tuning of $\tilde
m_{23}^2/\tilde m^2\cong 1$ to ${\cal O} (10^{-2})$ level
\cite{ArhribPRD01}. In fact, the scenario of our previous work
\cite{CHNPRL04} used $\tilde m_{22}^2=\tilde m_{33}^2=\tilde m^2$
and $1-\tilde m_{23}^2/\tilde m^2=0.04$ $(0.01)$ for $\tilde m=1$
$(2)$ TeV. The $\phi K_S$ sign anomaly in 2003 seemed to demand
such fine-tuning for light $\widetilde{sb}_1$. Furthermore, we
also demanded $m_{\tilde g}$ to be not heavier than 600 GeV. With
the weakening of the current ${\cal S}_{\phi K_S}$ data, the
previous somewhat extreme model parameters can be relaxed.
%We need to renew our understanding.
In the following, we continue to
pursue near maximal $\tilde s_R$-$\tilde b_R$ mixing for
convenience, {\it i.e.} $\theta=45^\circ$, but make the tuning
$1-\tilde m_{23}^2/\tilde m^2$ less strict.

Let us not give the explicit expressions~\cite{Chua01} of
gluino-quark-squark and squark-squark-gluino interactions in mass
eigenbasis. Instead, we present the right-handed chromo-dipole
contributions as an example. The relevant coefficient is described
as
\begin{equation}
C_{8}^{\prime G} = -\frac{\sqrt{2}\pi\alpha_s}{G_F\lambda_t}
\Biggl\{ \frac{1}{6} {\cal F}_1(\tilde m_1,m_{\tilde g},\tilde m)
+\frac{3}{2} {\cal F}_2(\tilde m_1,m_{\tilde g},\tilde m)
\Biggr\}, \nonumber \\
\end{equation}
where
\begin{eqnarray}
\!\!\!\!\!\!\!\!
& &
{\cal F}_1(\tilde m_1,m_{\tilde g},\tilde m)=
-c_\theta s_\theta e^{-i\sigma}\biggl[
\frac{f(\tilde x_1)}{\tilde m_1^2}-\frac{f(\tilde x_2)}{\tilde m_2^2}
\biggr]
\nonumber \\
& &\hspace{7mm}+ \tilde m m_{\tilde g}\biggl\{
\frac{c_\theta(c_\theta-s_\theta e^{-i\sigma})}{\tilde m_1^2-\tilde m^2}
\left(\frac{g(\tilde x_1)}{\tilde m_1^2}
-\frac{g(\tilde x_0)}{\tilde m^2}\right)
\nonumber \\
& & \hspace{1.7cm} +
\frac{s_\theta(s_\theta+c_\theta e^{-i\sigma})}{\tilde m_2^2-\tilde m^2}
\left(\frac{g(\tilde x_2)}{\tilde m_2^2}
-\frac{g(\tilde x_0)}{\tilde m^2}\right)
\biggr\}, \nonumber \\
\label{eq:C12pr}
\end{eqnarray}
with $\tilde x_0=m_{\tilde g}^2/\tilde m^2$, $\tilde x_1=m_{\tilde
g}^2/\tilde m_1^2$, and $\tilde x_2=m_{\tilde g}^2/\tilde m_2^2$,
and ${\cal F}_2(\tilde m_1,m_{\tilde g}, \tilde m)$ is obtained by
the replacement $f(\tilde x_i)\to h(\tilde x_i)$ and $g(\tilde
x_i)\to j(\tilde x_i)$.  The expressions of $f(\tilde x_i)$,
$g(\tilde x_i)$, $h(\tilde x_i)$ and $j(\tilde x_i)$ in
Eq.~(\ref{eq:C12pr}) are given by
\begin{eqnarray}
f(\tilde x_i)&=&\frac{1}{12(\tilde x_i -1)^4} \left(1-6\tilde
x_i+3\tilde x_i^2+2 \tilde x_i^3 -6\tilde x_i^2\ln\tilde
x_i\right),
\nonumber \\
g(\tilde x_i)&=& \frac{1}{2(\tilde x_i -1)^3} \left(1-\tilde
x_i^2+ 2\tilde x_i\ln\tilde x_i\right),
\nonumber \\
h(\tilde x_i)&=&\frac{1}{12(\tilde x_i -1)^4} \left(2+3\tilde
x_i-6\tilde x_i^2+\tilde x_i^3 +6\tilde x_i\ln\tilde x_i\right),
\nonumber \\
j(\tilde x_i)&=& \frac{1}{2(\tilde x_i -1)^3} \left(-3+ 4\tilde
x_i-\tilde x_i^2- \ln\tilde x_i\right).
\end{eqnarray}
The coefficients at $m_b$ scale can be calculated by means of the
leading order renormalization equations~\cite{Chua01},
\begin{eqnarray}
& &
C_{8}^{\prime G}(\mu)= \nonumber \\
& & \hspace{7mm}
\biggl(\frac{\alpha_s(m_t)}{\alpha_s(\mu)}\biggr)^{\frac{14}{23}}
\biggl(\frac{\alpha_s(\sqrt{M_{\rm SUSY}})}
{\alpha_s(m_t)}\biggr)^{\frac{14}{21}}
C_{8}^{\prime G}(\sqrt{M_{\rm SUSY}}), \nonumber \\
\label{eq:URNEW}
\end{eqnarray}
where we assume $M_{\rm SUSY}=\sqrt{\tilde m m_{\tilde g}}$.

Right-handed strong penguins $a^\prime_{3{\rm -}6}$ are also taken
into account in our calculation, where $a^\prime_{3,4}$ and
$a^\prime_{5,6}$ mainly enter $RR$ and $RL$ squark mixings,
respectively. The size of $a^\prime_{3{\rm -}6}$ is smaller than
the size of $C_{8}^{\prime G}$ by at least two orders of
magnitude.

Before moving to the numerical study in the next session, one
remark should be made. In general, the new effect on left-handed
dynamics arise from $LL$ and $LR$ squark mixing. Here we assume
that left-handed squark $\tilde q_L$ is at TeV scale. In addition,
one finds \cite{ArhribPRD01} that $\tilde s_L$-$\tilde b_L$ mixing
and $\tilde s_L$-$\widetilde{sb}_{1,2}$ mixing are suppressed by a
factor ${\cal O} (\lambda^2)$. Therefore, we expect the dominance
of the SM contribution in left-handed dynamics, and neglect the
left-handed SUSY effects. It is important to note that, in this
specific framework, a light $\widetilde{sb}_1$ can well survive
the constraint of $b\to s\gamma$ decay rate \cite{ArhribPRD01}. As
shown in Refs.~\cite{ArhribPRD01,CHNPRL04}, for $m_{\tilde g}>800$
GeV, neither the $\widetilde{sb}_1$ mass nor the $\sigma$ phase
are constrained. Furthermore, the impact of $\widetilde{sb}_{1}$
on the direct CPV in $b\to s\gamma$ \cite{bsphotonACP}, ${\cal
A}_{b\to s\gamma}$, is rather small. This is based on the fact
that there is no right-handed tree level weak interaction, and no
operator mixing between right- and left-handed dynamics. We found
that, even though ${\cal S}_{\phi K_S}<0$ had persisted, the
modulation from the SM prediction for ${\cal A}_{b\to s\gamma}$
(around +0.6\%) is as small as $\pm 0.15\%$.

%It is important to note that
%a light $\widetilde{sb}_1$ can well survive
%the constraint of $b\to s\gamma$ decay rate \cite{ArhribPRD01}.
%As shown in Refs.~\cite{ArhribPRD01,CHNPRL04}, for
%$m_{\tilde g}>800$ GeV, neither $\widetilde{sb}_1$ nor $\sigma$
%are constrained.
%The second remark is about SUSY contributions in our model to
%the left-handed dynamics. In general,
%the new effect on left-handed dynamics arise from
%$LL$ and $LR$ squark mixing.
%Here we assume that left-handed squark $\tilde q_L$ is at TeV scale.
%In addition, one finds \cite{ArhribPRD01} that
%$\tilde s_L$-$\tilde b_L$ mixing and
%$\tilde s_L$-$\widetilde{sb}_1$ mixing
%are suppressed by a factor $O(\lambda^2)$.
%Therefore, we expect the dominance of the SM contribution
%in the left-handed dynamics.
%In what follows, we will neglect the left-handed SUSY effects.

%%%%%%%% %%%%%%% %%%%%%% %%%%%%%
\section{\label{sec:result} \boldmath
Impact of $\widetilde{sb}_R$ on $\Delta{\cal S}_{f}$}

We now introduce the SUSY effects into $B_d\to \phi K^0$ and
$\pi^0 K^0$. As stated in previous sections, for simplicity we
neglect the right-handed SM contributions to $b\to s$ transition,
as well as the left-handed SUSY contributions. Consequently, in
our calculation, the left-handed coefficients $C_i$ are identical
to the SM calculation, while the right-handed coefficients
$C^\prime_i$ arise purely from NP.

%
% Fig.1
%
\begin{figure}[tbp]
  \centering
  \includegraphics[width=0.4\textwidth]{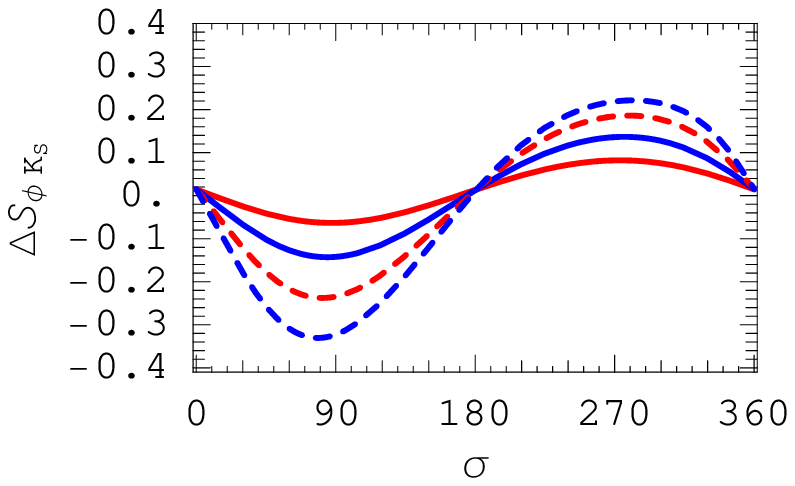}
  \includegraphics[width=0.4\textwidth]{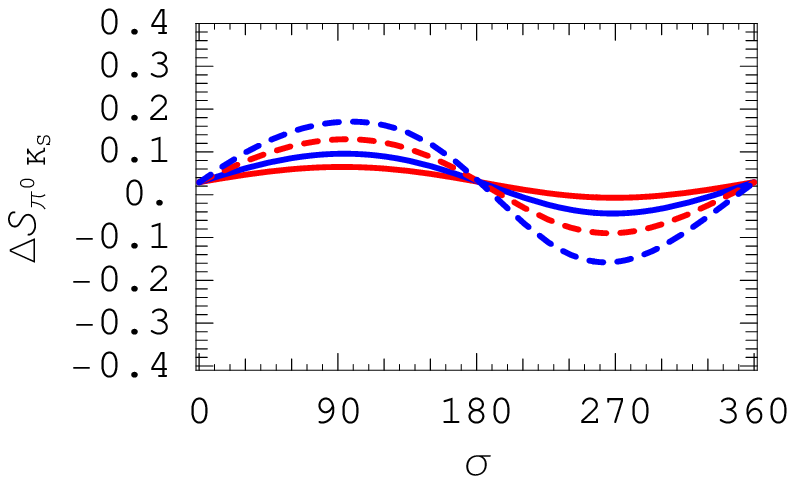}
  \caption{
 (a) $\Delta{\cal S}_{\phi K_S}$ and
 (b) $\Delta{\cal S}_{\pi^0 K_S}$ vs new $CP$ phase $\sigma$
 for $\tilde m=1.5$ TeV. Solid curves are for $m_{\widetilde{sb}_1}= 1100$ GeV
 and $750$ GeV with $m_{\tilde g}$ fixed at 950 GeV. Dashed curves are for
 $m_{\tilde g}=900$ GeV and $750$ GeV with $m_{\widetilde{sb}_1}$ fixed at 500 GeV.
 Steeper solid (dashed) curve corresponds to
 $m_{\widetilde{sb}_1}$ ($m_{\tilde g}$) $= 750$ GeV.}
  \label{fig:Fig1}
\end{figure}

The decay amplitude of $\bar B^0\to \phi \bar K^0$ in Eq.~(\ref{eq:ampphik})
is modified as
\begin{eqnarray}
\frac{{\cal M}(\phi \bar K^0)}{\epsilon^\ast\cdot P_B} &\propto&
\Biggl\{ \cdots + (C^{G}_{8}+C^{\prime G}_{8})
\frac{\alpha_s}{4\pi}\frac{m_b^2}{q^2}\tilde{S}_{\phi K} \Biggr\},
\label{eq:ampphiKNP}
\end{eqnarray}
where $\cdots$ are the terms shown in Eq.~(\ref{eq:ampphik}),
modified by $a_i \to a_i + a^\prime_i$. Because of $\widetilde
S_{\phi K}\sim {\cal O} (1)$ and $a_{3{\rm
-}6}^{\prime}/C_8^{\prime G} < {\cal O} (10^{-2})$, the
chromo-dipole SUSY effect is the dominant NP effect in
Eq.~(\ref{eq:ampphiKNP}).

Fig.~\ref{fig:Fig1}(a) illustrates our result for $\Delta{\cal
S}_{\phi K_S}$ vs new CP phase $\sigma$. We fix $\tilde m$ to be
1.5 TeV for illustration. Since NF contains small strong phases,
our results reflect genuine NP contributions. Here we set
$\Delta{\cal S}_{\phi K_S}\simeq -0.2$ as a target for extracting
the preferred regions for $m_{\widetilde{sb}_1}$ and $m_{\tilde
g}$. We illustrate with four examples spanning some range for
$m_{\widetilde{sb}_1}$ and $m_{\tilde g}$, but with values
remaining {\it natural}. We see that, for $\sigma\sim 90^\circ$,
the range $m_{\widetilde{sb}_1}\simeq 500$--$800$ GeV with
$m_{\tilde g}\simeq 700$--$900$ GeV can be easily accommodated by
the current $\Delta{\cal S}_{\phi K_S}$ data. These mass regions
are well within the discovery range at LHC~\cite{CMSBrochure},
which would be commissioned in 2007. On the other hand, one has
$\Delta{\cal S}_{\phi K_S}\gg \Delta{\cal S}_{\phi K_S}^{\rm
SM}>0$ for $\sigma \sim 270^\circ$, and $\Delta{\cal S}_{\phi
K_S}\simeq \Delta{\cal S}_{\phi K_S}^{\rm SM}$ at $\sigma \simeq
0^\circ/180^\circ$.

Note that $m_{\widetilde{sb}_1}\ll m_{\tilde g}\sim \tilde m$, or
$m_{\tilde g}\ll m_{\widetilde{sb}_1}\sim\tilde m$ as well as
$m_{\widetilde{sb}_1}\sim m_{\tilde g}\ll\tilde m$ with small
$\sigma$ phase \cite{CHNPRL04} can still be accommodated and need
not be thrown away. However, these ranges are not of interests in
this work as we emphasize naturalness for $m_{\widetilde{sb}_1}$
on the SUSY scale. In the following, we take
$[m_{\widetilde{sb}_1},m_{\tilde g}]\simeq[500,900]$ GeV as our
standard value with $\tilde m$ at 1.5 TeV.

The decay amplitude of $\bar B^0\to \pi^0 \bar K^0$
is modified as
\begin{eqnarray}
{\cal M}(\pi^0 \bar K^0) &\propto& \Biggl\{ \cdots +
(C^{G}_{8}-C^{\prime G}_{8})
\frac{\alpha_s}{4\pi}\frac{m_b^2}{q^2}\tilde{S}_{\pi K} \Biggr\},
\label{eq:amppiKNP}
\end{eqnarray}
where $\cdots$ are the terms shown in Eq.~(\ref{eq:amppik}) with
replacement $a_i \to a_i - a^\prime_i$. Contrary to
Eq.~(\ref{eq:ampphiKNP}), the primed terms change sign, which
would cause anticorrelation effect between ${\cal S}_{\phi K_S}$
and ${\cal S}_{\pi^0 K_S}$ \cite{CHNPRL04}.
Fig.~\ref{fig:Fig1}(b) illustrates our result of $\Delta{\cal
S}_{\pi^0 K_S}$ which shows the opposite trend to $\Delta{\cal
S}_{\phi K_S}$. Like $\phi K$ amplitude, the dominant SUSY effect
would come from the chromo-dipole term. However, the new effect on
$\Delta{\cal S}_{\pi^0 K_S}$ is milder since the chiral operator
$O_{6}$ enhances the left-handed, {\it i.e.} the SM,
contributions. Specifically, our standard value gives $\Delta{\cal
S}_{\pi^0 K_S}\simeq 0.13$ for $\sigma\sim 90^\circ$, while
$\Delta{\cal S}_{\pi^0 K_S}\simeq -0.09$ for $\sigma\sim
270^\circ$.

For the current data of $\Delta{\cal S}_{\phi K_S}\simeq -0.22\pm
0.19$, the accuracy is not sufficient to discriminate between the
NP result and the SM result.
By continuing to run until 2008, the data accumulated by the B
factories is expected to be more than $3\times 10^9$ $B\bar B$
events, about four times the current number, which can reduce the
current error of $\Delta{\cal S}_{\phi K_S}$ by
half~\cite{HazumiICFP05}. In order for an order of magnitude
improvement in accuracy, we may need to wait for the {\it Super B
factory}~\cite{HazumiICFP05}. Bearing  this in mind, let us think
about future prospects along the four scenarios enumerated in the
Introduction. For sake of discussion, we split $\sigma$ into 4
regions: \\
\indent\hskip0.2cm (a) $90^\circ\pm 60^\circ$, \\
\indent\hskip0.2cm (b) $270^\circ\pm 60^\circ$, \\
\indent\hskip0.2cm (c) $0^\circ\pm 30^\circ$, and \\
\indent\hskip0.2cm (d) $180^\circ\pm 30^\circ$.

Because of the anticorrelation between $\Delta{\cal S}_{\phi K_S}$
and $\Delta{\cal S}_{\pi^0 K_S}$, one has $\Delta{\cal S}_{\pi^0
K_S}>0$ for $\Delta{\cal S}_{\phi K_S}<0$, and vice versa for
$\widetilde{sb}_R$ effect. We see that {\it Scenario 1} prefers
the $\sigma$ range (a), which would demand $\Delta{\cal S}_{\pi^0
K_S}$ to be positive, the same as the SM prediction, and likely
larger.
%, which are inconsistent with the
%present data of ${\cal S}_{\pi^0 K_S}$.
If the present $\Delta{\cal S}_{\phi K_S}$ holds, but $\Delta{\cal
S}_{\pi^0 K_S}$ changes sign in the future, the range (a) with
$m_{\tilde g}\lesssim 900$ GeV might be the solution. {\it
Scenario 2}, instead, would require the further experimental
results to raise up $\Delta{\cal S}_{\phi K_S}$, but confirm the
present sign of $\Delta{\cal S}_{\pi^0 K_S}$. This scenario, if
realized, would prefer range (b). {\it Scenario 3} is consistent
with the current experimental results of $\Delta{\cal S}_{\phi
K_S,\pi^0 K_S}<0$. However, our model, as well as any SUSY model
with underlying large right-handed dynamics, cannot accommodate
this scenario because of the implied anticorrelations. If the
present deviations in $\Delta{\cal S}_{\phi K_S,\pi^0 K_S}$
persist, it would evidently be a hint for NP, but would be a
different picture of NP than the current one (a study within a
certain specific model will be described in Ref.~\cite{SM4DeltaS})
studied here. The ranges (c) and (d) can fit {\it Scenario 4}
which implies that $\Delta{\cal S}_{\phi K_S}$ and $\Delta{\cal
S}_{\pi^0 K_S}$ will converge to $\sin 2\phi_1/\beta$ in the
future. In this scenario, we cannot distinguish between NP and SM
since the effect from NP might overlap with the effect from SM.
However,in the next section, we shall see that measurements of
$B_s$ mixing and the associated CPV phase may discover NP, even if
{\it Scenario 4} is realized.

%From straightforward calculations, we evaluated
%our SUSY effects on TCPVs in
%$\eta^\prime K_S$, $\omega K_S$, $\rho^0 K_S$.
%We found, as noted in Refs.~\cite{rhSUSYKagan,rhSUSYKhalil},
%$\Delta{\cal S}_{\eta^\prime K_S}\simeq
%\Delta{\cal S}_{\pi^0 K_S}$.
%Our results of $\Delta{\cal S}_{\omega K_S}$ and
%$\Delta{\cal S}_{\rho K_S}$ show the opposite trend
%with $\Delta{\cal S}_{\phi K_S}$.
%These results depend strongly on the form factors
%since the hadoronic parameters
%$\widetilde S_{\omega K}$ and $\widetilde S_{\rho K}$
%come from $B\to\omega,\rho$ transitions, and
%are quite sensitive to the form factors.
%For our results of TCPVs in
%$\pi^0 K^\ast$ and $\eta^\prime K_S$,
%we found sizable NP effects in $\pi^0 K^\ast$, while
%invisibly small in $\eta^\prime K^\ast$.
%
%
%Since NF contains small strong phases,
%our results reflect real NP contributions.
%However we need to caution that the hadronic factor
%$\tilde S_f/q^2$ brings large uncertainties into our results.
%Furthermore, more practical application of factorization framework
%could dilute the NP effects on $\Delta{\cal S}_f$, especially
%for $\eta^\prime K_S$, $\omega K_S$ and $\rho^0 K_S$ because
%large corrections needed to reach observed rate.

From straightforward calculations, we have also evaluated our SUSY
model effects on $\Delta{\cal S}_{f}$ for  $f = \eta^\prime K_S$,
$\omega K_S$, $\rho^0 K_S$, $\pi^0 K^\ast$ and $\eta^\prime
K^\ast$. So far, ${\cal S}_{\eta^\prime K_S}$ and ${\cal
S}_{\omega K_S}$ have been measured. For $\omega K_S$, the BaBar
result \cite{BabaromegaK} and Belle result \cite{Bellebsss} are
not in good agreement, as is the case for ${\cal S}_{\eta^\prime
K_S}$. Recently, BaBar measured ${\cal S}_{\pi^0\pi^0K_S}$
\cite{BabarpipiK} which is, in principle, equivalent to ${\cal
S}_{\pi^0 K^\ast}$, where $K^\ast\to \pi^0K_S$, however the error
is very large.

We confirm that the effect on $\Delta{\cal S}_{\eta^\prime K_S}$
anticorrelates \cite{rhSUSYKagan,rhSUSYKhalil} with the effect on
$\Delta{\cal S}_{\phi K_S}$. Our result of $\Delta{\cal
S}_{\eta^\prime K_S}$ is rather similar to $\Delta{\cal S}_{\pi^0
K_S}$. On the other hand, $\Delta{\cal S}_{\omega K_S}$ and
$\Delta{\cal S}_{\rho^0 K_S}$ show the opposite trend with
$\Delta{\cal S}_{\phi K_S}$. The hadronic parameters $\widetilde
S_f$ in $\eta^\prime K_S$, $\omega K_S$ and $\rho^0 K_S$ are more
involved than $\widetilde S_{\phi K_S}$ or $\widetilde S_{\pi^0
K_S}$. In particular, our results for $\omega K_S$ and $\rho^0
K_S$ are quite sensitive to form factors. This is because the
hadronic parameters in these modes arise from $B\to V$
transitions, unlike $\phi K_S$, and depend strongly on the form
factors (see Eq.~(\ref{eq:tilSBV})). Within NF, our SUSY effects
on $\Delta{\cal S}_{\eta^\prime K_S}$, $\Delta{\cal S}_{\omega
K_S}$ and $\Delta{\cal S}_{\rho^0 K_S}$ could be measurable.
However,
%we need to caution that
the large rates for these modes are not well understood. Our SUSY
effects do not help to reach the observed rates. The additional
hadronic effects needed to enhance the rates could likely be
rather large and smear the impact of our SUSY effects very much.
This is why we refrain from going into the details about these
decays. For our results on TCPV in $\pi^0 K^\ast$ and $\eta^\prime
K^\ast$, we just mention that we found sizable NP effects in
$\pi^0 K^\ast$ with variation around the SM result, while
invisibly small for $\eta^\prime K^\ast$.

%
%Super B factory:\cite{HazumiICFP05}
%The accuracy of the current data $\Delta{\cal S}_{\phi K_S}\simeq -0.22\pm 0.19$
%is not sufficient to discriminate between NP results and SM prediction.
%It is expected that half an error in current $\Delta{\cal S}_{\phi K_S}$
%is reduced at currently running B factories within 5 years.
%Moreover, Super B factory, if comes true,
%can ensure the present error to 1/10 by accumulating data for 10 years.

%%%%%%%% %%%%%%% %%%%%%% %%%%%%%
\section{\label{sec:better} \boldmath
The Probes $\Delta m_{B_s}$, $\sin2\Phi_{B_s}$ and ${\cal
S}_{K^\ast\gamma}$}

Due to the softening of TCPV data on $\Delta{\cal S}_{\phi K_S}$
and $\Delta{\cal S}_{\pi^0 K_S}$, we find that the preferred
regions for $m_{\widetilde{sb}_1}$ and $m_{\tilde g}$ are now
heavier hence more natural. However, the parameter $\widetilde
S_f/q^2$ related to the hadronic matrix element of chromo-dipole
operator is highly uncertain. Furthermore, annihilation phases are
likely present in these hadronic $B$ decays. Such hadronic
uncertainties would prevent the determination of the SUSY model
parameters, even when $\Delta{\cal S}_{\phi K_S}$ and $\Delta{\cal
S}_{\pi^0 K_S}$ become precisely measured. It is of interest to
explore processes that are largely free of hadronic uncertainties.

As pointed out in Ref.~\cite{ChuaPRD99}, $\Lambda$ polarization in
$\Lambda_b\to \Lambda\gamma$ \cite{Lambdapol} would be a clean
probe of the right-handed NP. However, the impact from our updated
heavier $m_{\widetilde{sb}_1}$ and $m_{\tilde g}$ would give rise
to effects no larger than 10\%. Besides, this mode seems very hard
to measure. In this sense, $B_s$ mixing and TCPV in $B_d\to K^\ast
\gamma$ would be better probes for our SUSY effects
\cite{ArhribPRD01,CHNPRL04}. We therefore update these measurables
to the current preferred parameter space for our model.

\subsection{\boldmath $B_s$ system}

%
% Fig.2
%
\begin{figure}[tbp]
  \centering
  \includegraphics[width=0.403\textwidth]{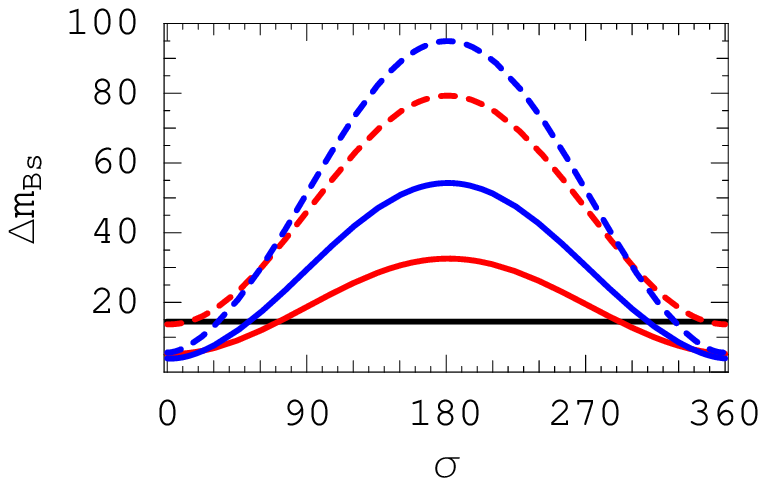}
  \includegraphics[width=0.4\textwidth]{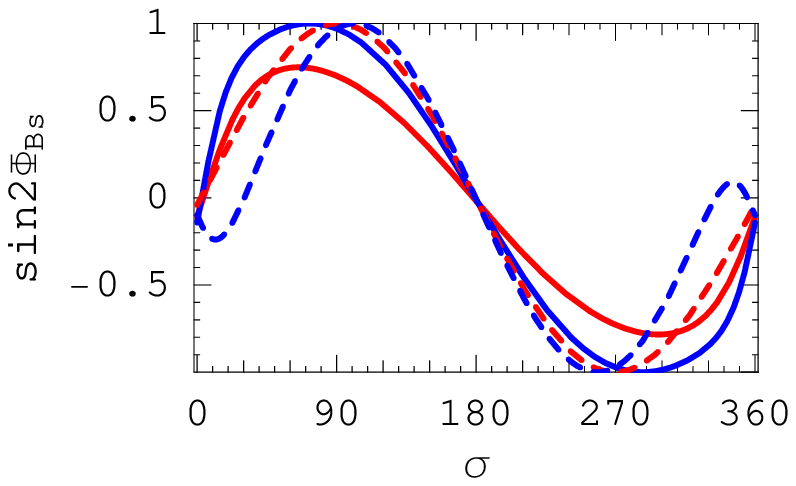}
  \caption{
  (a) $\Delta m_{B_s}$ and (b) $\sin 2\Phi_{B_s}$ vs
  $\sigma$ for $\tilde m=1.5$ TeV. Notation is the same as Fig.~\ref{fig:Fig1}.
  The horizontal line in (a) is the current $\Delta m_{B_s}$ bound.}
  \label{fig:Fig2}
\end{figure}

$B_s$ mixing can be studied by flavor specific decays,
such as $B_s\to D_s^+\pi^-$.
The present bound of the mass difference is
$\Delta m_{B_s}>14.5$ ps$^{-1}$ with 95\% confidence level \cite{PDG}.
The SM prediction of $\Delta m_{B_s}$, on the other hand,
is around 20 ps$^{-1}$.
%Measurement of $B_s$ mass difference $\Delta m_{B_s}$ is
%one of the prime target at currently running Tevatron.
Uncovering a $\Delta m_{B_s}$ value considerably larger than this
would be a hint for NP. The currently running Tevatron can cover
part of the SM range. But at the LHC, besides the collider
detectors ATLAS and CMS, the dedicated LHCb experiment claims to
cover $\Delta m_{B_s}$ up to 70 ps$^{-1}$, and should have no
problem in fully exploring $\Delta m_{B_s}^{\rm SM}$.

In our previous work, we studied the implication of $\phi K_S$
sign anomaly for $\Delta m_{B_s}$ \cite{CHNPRL04}. The negative
${\cal S}_{\phi K_S}$ would imply that $B_s$ probably oscillates
faster than $70$~ps$^{-1}$, which would be challenging even for
LHCb. The situation is now relaxed. It is useful to revisit the
impact of our SUSY effect on $\Delta m_{B_s}$.

The error on theoretical calculations are contained in the
hadronic factor $f_{B_s}\sqrt{B_{B_s}}$, which is being studied
vigorously by lattice approaches. There is no strong phase,
therefore $\Delta m_{B_s}$ should eventually supply much cleaner
information about new CPV phase.

Fig.~\ref{fig:Fig2}(a) shows our prediction of $\Delta m_{B_s}$.
%Our SM calculation leads to
%$\Delta m_{B_s}^{\rm SM}\sim 16$ ps$^{-1}$.
%We take $\sqrt{B_{B_s}}f_{B_s}=240$ MeV.
We plot the present bound as horizontal solid straight line. Let
us focus on our standard value $[m_{\widetilde{sb}_1},m_{\tilde
g}]\simeq[500,900]$ GeV. For the $\sigma$ range (a) and (b),
$\Delta m_{B_s}$ lie in the measurable $20$-$70$ ps$^{-1}$ range.
The curious {\it Scenario 4} prefers the $\sigma$ ranges (c) and
(d). From $\Delta {\cal S}_{f}$ studies, one cannot distinguish
these two possibilities, therefore cannot tell whether one has NP
or not. Interestingly, $\Delta m_{B_s}$ can disclose the
remarkable difference between the ranges (c) and (d). For the
range (c), $\Delta m_{B_s}$ is similar to the SM prediction. For
the range (d), instead, one has $\Delta m_{B_s}\sim 70$-$80$
ps$^{-1}$.

Assuming $\Delta m_{B_s}$ can be observed, the associated CPV
phase $\sin 2\Phi_{B_s}$ can provide further information on
$\sigma$. $\sin 2\Phi_{B_s}$ can be measured for example via
$B_s\to J/\psi \Phi$ decay. Similar to the $B_d$ case, this mode
is dominantly a $b\to c\bar cs$ transition, hence probes the weak
phase associated with $B_s$ mixing. The SM asserts $\sin
2\Phi_{B_s}\sim -0.04$. More than $\pm 0.1$ deviation from the SM
prediction, if measured, would indicate the presence of new CPV
phase.
As shown in Fig.~\ref{fig:Fig2}(b), the various heavier
$m_{\widetilde{sb}_1}$ and $m_{\tilde g}$ values can still make
spectacular impact on $\sin 2\Phi_{B_s}$. Note that {\it Scenario
1} implies $\sin 2\Phi_{B_s}\sim +1$, while {\it Scenario 2}
implies $\sin 2\Phi_{B_s}\sim -1$. Measurement of $\sin
2\Phi_{B_s}$ and $\Delta m_{B_s}$ might provide confirmation of
{\it Scenario 1}, {\it 2} or {\it 4}.

Besides $\sin 2\Phi_{B_s}$ measurement, one can study TCPV in
nonleptonic $B_s$ decays as well. For $b\to s$ transition, the
promising decays are $B_s\to K^+K^-$
\cite{Fleischer,HNPRD04,Beak05} and $\phi\phi$, while measuring
$K^0\bar K^0$ might be hopeless. Especially, $B_s\to\phi\phi$ will
be a help to understand the polarization anomaly in $B_d\to\phi
K^\ast$. The SM predicts ${\cal S}_{B_s \to K^+K^-}\simeq 0.5$
reflecting its decay phase from $V_{us}^\ast V_{ub}$ accompanied
by $a_1$ in Table~\ref{tab:Wilson}, while ${\cal S}_{B_s \to
\phi\phi}\simeq 0$ due to the absence of the decay phase just as
in $B_d\to \phi K_S$.
%The former reflects its decay phase from $V_{us}^\ast V_{ub}$
%accompanied by $a_1$ in Table~\ref{tab:Wilson},
%while the decay phase in $\phi\phi$
%is approximately zero from the same reason with $B_d\to \phi K_S$.
We find that ${\cal S}_{B_s \to K^+K^-}$ and ${\cal S}_{B_s \to
\phi\phi}$ (longitudinal polarization state) show the opposite
trend with $\sin 2\Phi_{B_s}$. Finding ${\cal S}_f$ in $B_s$
decays would be a further crosscheck of hint for NP from
corresponding $B_d$ decays.

\subsection{\boldmath TCPV in $B_d\to K^\ast\gamma$}

TCPV in $B\to K^\ast \gamma$ is also a good place to look for SUSY
effect in our model \cite{CHNPRL04}.
%Here we revisit the implication for ${\cal S}_{K^\ast \gamma}$.

The photon radiated from $b\to s\gamma$ can, in principle, pick up
the chirality information of the electromagnetic operators arising
from the left- and right-handed dynamics. Therefore, the SM
contribution dominantly generates the left-handed polarized
photon, while our SUSY effect dominantly generates the
right-handed photon. Although ${\cal S}_{K^\ast \gamma}$ is
studied without directly measuring the photon handedness, because
of the requirement that the final state should be a $CP$
eigenstate, both components need to be present for ${\cal
S}_{K^\ast \gamma}$ to be nonvanishing. Summing over the two
separate rates for $K^\ast\gamma_L$ and $K^\ast\gamma_R$, the $CP$
violating interference can occur. Following
Refs.~\cite{AGS,ChuaPRD99}, one has
\begin{eqnarray}
{\cal S}_{K^\ast\gamma}= \frac{2{\rm
Im[e^{-i2\phi_1}C_7^{\gamma}C_7^{\prime\gamma}]}}
{\left|C_7^\gamma\right|^2+\left|C_7^{\prime\gamma}\right|^2},
\label{eq:SKphoto}
\end{eqnarray}
where we follow the convention of Eq.~(\ref{eq:AfSf}) which is
opposite to that used in Ref.~\cite{AGS}. In the SM,
$C_7^{\prime\gamma}/C_7^{\gamma}\simeq m_s/m_b$, hence ${\cal
S}_{K^\ast \gamma}\simeq -0.04$ is predicted.

%
% Fig.3
%
\begin{figure}[tbp]
  \centering
  \includegraphics[width=0.403\textwidth]{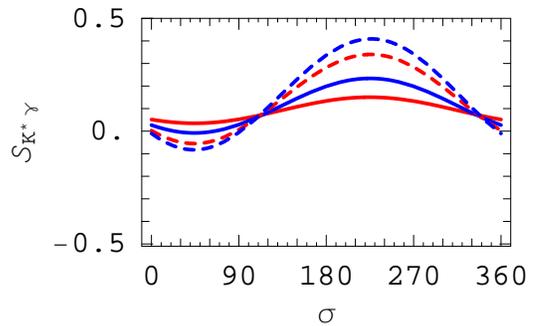}
  \caption{
  ${\cal S}_{K^\ast \gamma}$ vs
  $\sigma$ for $\tilde m=1.5$ TeV.
  Notation is the same as Fig.~\ref{fig:Fig1}.}
  \label{fig:Fig3}
\end{figure}

Fig.~\ref{fig:Fig3} illustrates our prediction with our updated
$m_{\widetilde{sb}_1}$ and $m_{\tilde g}$, which shows a
measurable effect~\cite{note}. We see that {\it Scenario 1} gives
small ${\cal S}_{K^\ast \gamma}$ with negative sign, while sizable
${\cal S}_{K^\ast \gamma}$ with positive sign will be realized in
{\it Scenario 2}. Measurement of ${\cal S}_{K^\ast \gamma}$ can
also help discriminate between the $\sigma$ range (c) and (d). So
far, errors in the experimental results at BaBar
\cite{BabarKphoto} and at Belle \cite{BelleKphoto} are too large
to observe a real hint for NP.
%We may need to wait for the next generation B factory.

We make one cautionary remark. ${\cal S}_{K^\ast \gamma}$ has been
expected to be largely free from hadronic uncertainties. A recent
study of TCPV in $B\to X\gamma$ \cite{GrinsteinPRD05} in Soft
Collinear Effective Theory claims that the SM prediction of ${\cal
S}_{K^\ast \gamma}$ could easily be of order 0.1 (the sign cannot
be fixed) with large uncertainties from unknown hadronic matrix
elements. If such sizable uncertainties are indeed present in SM,
${\cal S}_{K^\ast \gamma}$ may not be as good a probe of NP as
$\sin 2\Phi_{B_s}$. Even so, however, improved measurements in the
future will test our result for ${\cal S}_{K^\ast \gamma}$ if
$\sigma \in (180^\circ, 300^\circ)$, i.e. the ranges (b) and (d).

%\item comment on DCPV in $b\to s\gamma$
%DCPV in $b\to s\gamma$ can be accessible \cite{bsphotonACP}.
%Neglecting the left-handed SUSY contributions, the deviation from the SM prediction
%comes from
%\begin{eqnarray}
%{\cal A}^{(R)}_{\rm{CP}}=
%\frac{a_{87}{\rm Im}[C_8^{G\prime}C_7^{\gamma\prime\ast}]}
%{\left|C_7^\gamma\right|^2+\left|C_7^{\gamma\prime}\right|^2}
%\end{eqnarray}
%${\cal A}^{\rm SM}_{\rm{CP}}=$
%We find negligibly small modulation (smaller than 0.05\%).
%\end{itemize}

\section{\label{sec:Epi}Discussion and Conclusion}

%\begin{eqnarray}
%\tilde\alpha=
%\frac{\left|C_7^\gamma\right|^2-\left|C_7^{\gamma\prime}\right|^2}
%{\left|C_7^\gamma\right|^2+\left|C_7^{\gamma\prime}\right|^2}
%\end{eqnarray}

%NF is not sufficient to account for the decay rates.
As stated previously, a more up to date application of the
factorization framework could dilute the NP effects on
$\Delta{\cal S}_f$ \cite{SM4DeltaS}. Let us try to estimate the
probable dilution effect on $\Delta{\cal S}_{\phi K_S}$. To
address this, we pay attention to the decay rate, and introduce a
heuristic term for annihilation effect, by referring to QCDF
framework \cite{QCDF}. Without the chromo-dipole term, multiplying
$a_i + a^\prime_i$ in the SM by $(1+0.4 \, e^{i 30^\circ})$
enhances the decay rate from ${\cal B}\simeq 5.3\times 10^{-6}$ to
$9.8\times 10^{-6}$. The direct CPV around $+0.02$ is not in
disagreement with data. Adding the chromo-dipole term (but not
scaled by the annihilation factor), our standard values for NP
parameters give ${\cal B}\simeq 7.6\times 10^{-6}$ for $\sigma\sim
90^\circ$, which remains within the experimental error. In this
case, NP effect on TCPV gets diluted by the annihilation effect,
and $\Delta{\cal S}_{\phi K_S}$ becomes 35\% weaker than shown in
Fig.~\ref{fig:Fig1}(a). Because of this dilution from non-$CP$
related physics, and because the associated strong phase is not
very large, the effect on direct CPV is small and still consistent
with data.
%gets enhanced only to $+0.03$ for
%$\sigma\sim 90^\circ$.
%Setting $\tilde S_{\phi K}=0$, $a_i +
%a^\prime_i$ multiplied by $(1+0.3 e^{i 40^\circ})$ gives
%enhancement for the decay rate from ${\cal B}\simeq 5.3\times
%10^{-6}$ to $8.3\times 10^{-6}$, and direct CPV around $+0.02$
%which agrees with data. In this case, NP effect gets diluted and
%$\Delta{\cal S}_{\phi K_S}$ becomes 27\% weaker than
%Fig.~\ref{fig:Fig1}(a). Because of this dilution from non-$CP$
%related physics, and because the associated strong phase is not
%very large, the direct CPV gets enhanced only to $+0.04$ for
%$\sigma\sim 90^\circ$ and remains within error.
From a similar study, we find that the dilution for $\Delta{\cal
S}_{\pi^0 K_S}$ could be 42\%. We therefore caution that NP effect
in TCPV observables might get diluted by such hadronic effects.

As our model still adheres to large right-handed dynamics, we
should comment on the constraint from EDM of $^{199}{\rm H_g}$. As
claimed in Ref.~\cite{HisanoPLB04}, within NP CPV with underlying
SUSY, the chromoelectric dipole moment (CEDM) of $s$ quark
($d^C_s$) would strongly correlate with ${\cal S}_{\phi K_S}$.
%$d^C_s$ is constrained by the EDM of $^{199}{\rm H_g}$, and
%current bound is $|d_s^c|<5.8\times 10^{-25}$ cm.

In our model, $d^C_s$ receives contributions from $RR$ squark
mixing as well as $RL$ squark mixing, where the latter is enhanced
by $m_b/m_s$ %compared with $b\to s$ transition,
and accompanied with $\tilde s_L$-$\tilde{b}_{L}$ mixing parameter
$(\delta_{LL})_{sb}$. Even if we set $(\delta_{LL})_{sb}=0$,
$d^C_s$ for our standard value turns out to be three times larger
than the present bound \cite{HisanoPLB04}. However, there could be
cancellations between effect from $RR$ mixing and from $RL$
mixing. As a particular case, taking $(\delta_{LL})_{sb}\simeq
-0.01$ would allow our updated $m_{\widetilde{sb}_1}$ and
$m_{\tilde g}$ values. The left-handed squark mixing in our model
should therefore be studied further.

%\begin{eqnarray}
%& &\!\! \tilde d^{c} = \frac{\alpha_s m_s}{2\pi}
%\Biggl\{ -\frac{1}{6}
%{\rm Im}[{\cal\tilde F}_1(\tilde m_1,m_{\tilde g},\tilde m)]
%-\frac{3}{2}
%{\rm Im}[{\cal\tilde F}_2(\tilde m_1,m_{\tilde g},\tilde m)]
%\Biggr\}, \nonumber \\
%& & {\cal\tilde F}_1(\tilde m_1,m_{\tilde g},\tilde m)=
%-c_\theta s_\theta e^{-i\sigma}\biggl[
%\frac{f(\tilde x_1)}{\tilde m_1^2}-\frac{f(\tilde x_2)}{\tilde m_2^2}
%\biggr]
%\nonumber \\
%& &\hspace{7mm}+ \tilde m m_{\tilde g}(\delta_{LL})_{sb}
%\frac{m_b}{m_s}\biggl\{
%\frac{c_\theta(c_\theta-s_\theta e^{-i\sigma})}{\tilde m_1^2-\tilde m^2}
%\left(\frac{g(\tilde x_1)}{\tilde m_1^2}
%-\frac{g(\tilde x_0)}{\tilde m^2}\right)
%\nonumber \\
%& & \hspace{1.7cm} +
%\frac{s_\theta(s_\theta+c_\theta e^{-i\sigma})}{\tilde m_2^2-\tilde m^2}
%\left(\frac{g(\tilde x_2)}{\tilde m_2^2}
%-\frac{g(\tilde x_0)}{\tilde m^2}\right)
%\biggr\},
%\end{eqnarray}

Even if studies of CPV can determine $m_{\widetilde{sb}_1}$,
$m_{\tilde g}$ and $\sigma$, direct observation of
$\widetilde{sb}_{1}$ is much more exciting. A 200 GeV
$\widetilde{sb}_1$ as suggested by us when ${\cal S}_{\phi K_S}<0$
could be discovered at the Tevatron. However, with the weakening
of the current ${\cal S}_{\phi K_S}$ data,
$m_{\widetilde{sb}_1}\gtrsim 500$ GeV would be the more plausible
range. Observation of such $\widetilde{sb}_1$ is hopeless at the
Tevatron, but very relevant at the LHC.

Assuming that a bino $\tilde \chi_1^0$ is lighter than
$\widetilde{sb}_{1}$ \cite{ArhribPRD01}, the most interesting
decay mode is $\widetilde{sb}_1 \to b/s\;\tilde \chi_1^0$. Since
$\widetilde{sb}_1$ carries both $s$ and $b$ flavors, both the
$\widetilde{sb}_1\to b\tilde{\chi}^0$ and $\widetilde{sb}_1\to
s\tilde{\chi}^0$ decays, depending on the mixing angle
$\sin\theta$ of Eq. (\ref{eq:masseigen}), could be comparable in
rate. If maximal $\tilde s_R$-$\tilde b_R$ mixing is realized, the
standard $\tilde b$ squark search bound, based on $b$-tagging,
would be weakened.
We stress that this possibility should be kept in mind for direct
search. It would be exciting to discover a flavor-mixed squark, as
SUSY and flavor would become clearly linked to each other.
The production of light $\widetilde{sb}_1$ at hadron colliders
have been studied by paying attention to lower mass
$m_{\widetilde{sb}_1}$ and $m_{\tilde g}$ as hinted by ${\cal
S}_{\phi K_S}<0$ \cite{CheungPRD04}, which is aimed more at the
Tevatron. But the era of LHC is approaching. With the heavier
$m_{\widetilde{sb}_1}$ and $m_{\tilde g}$ range now implied by the
softened $\Delta {\cal S}_f$, the studies for collider search
should be updated.

Let us conclude. The hint for New Physics in $B_d\to\phi K_S$ has
weakened. The mild hint no longer calls for rather light
$m_{\widetilde{sb}_1}$ and $m_{\tilde g}$. Comparing with recent
experimental data, we extract the new mass range of
$m_{\widetilde{sb}_1}\simeq 500-800$ GeV and $m_{\tilde g}\simeq
700-900$ GeV, which is much more natural on the SUSY scale of
$\tilde m \sim 1-2$ TeV. $B_s$ mixing, the associated $CP$
violation $\sin 2\Phi_{B_s}$, and time-dependent $CP$ violation in
$B_d\to K^\ast \gamma$ would be better probes of our flavor/SUSY
model effects, which are less affected by hadronic uncertainties.
The impact on $\Delta m_{B_s}$ and $\sin 2\Phi_{B_s}$ is very
relevant to LHCb. By the time of LHC turn-on, we may know better
about the hints for NP from $\Delta{\cal S}_{f}$. But $B_s$
studies and direct search for the $\widetilde{sb}_1$ squark at the
LHC would open a new era.

%%%%%%%% %%%%%%% %%%%%%% %%%%%%%
%%%%%%%% %%%%%%% %%%%%%% %%%%%%%
\begin{acknowledgments}
This work is supported in part by grants NSC-94-2112-M-002-035 and
NSC-94-2811-M-002-053.
\end{acknowledgments}

%%%%%%%% %%%%%%% %%%%%%% %%%%%%%
%%%%%%%% %%%%%%% %%%%%%% %%%%%%%
\appendix
\begin{widetext}
\section{\label{app:tildeS} \boldmath
hadronic parameter $\widetilde S_{f}$}

We compute the hadronic parameter $\widetilde S_f$ of
chromo-dipole operator contribution to $B\to f$ decay by following
Ref.~\cite{ChromoDipole}. The expressions depend on $f$ being
$PP$, $VP$ or $VV$ final state, where $P$ is a pseudoscalar, and
$V$ a vector meson.

\subsection{$B\to PP$ decays}
With $B\to P_2$ transition, one gets
\begin{eqnarray}
\widetilde S_{P_1P_2}
 &=& -\frac{8}{9} \Biggl[\Biggl(1+\frac{2m_1^2}{(m_s+m_q)(m_b-m_q)} \Biggr)
                         \biggl(1-\frac{m_s}{m_b}\biggr)
                         + \frac{(P_B\cdot p_1)}{m_b(m_b-m_q)}
\nonumber \\
 & & \hskip0.9cm
 -\frac{m_1^2(2m_B^2-(P_B\cdot p_1))}{m_b(m_B^2-m_2^2)(m_s+m_q)}
 +\frac{m_B^2m_1^2-(P_B\cdot p_1)^2}{2m_bm_B(m_B^2-m_2^2)} \Biggr].
 \label{eq:tilSBP}
\end{eqnarray}

\subsection{$B\to VP$ decays}

For $B\to P$ transition, one has
\begin{eqnarray}
\widetilde S_{V_1P_2}= -\frac{8}{9}
\Biggl[
1+\frac{m_s}{m_b}+
\frac{m_B^2-m_2^2}{2m_b(m_b-m_q)}
-\frac{f_1^T}{f_1}\frac{m_1}{2m_b}
-\frac{(P_B\cdot p_1)}{4m_bm_B}
\Biggr],
\label{eq:tilSBPV}
\end{eqnarray}
where we assume $f^T_1=f_1$ in our calculation.

For $B\to V$ transition, one has
\begin{eqnarray}
\widetilde S_{P_1V_2}
 &=& -\frac{8}{9}
      \Biggl[\Biggl(1-\frac{2m_1^2}{(m_s+m_q)(m_b+m_q)}\Biggr)
             \biggl(1 + \frac{m_s}{m_b}\biggr)
             +\frac{(P_B\cdot p_1)}{m_b(m_b+m_q)}
\nonumber \\
 & & \hskip0.9cm
 -\frac{m_1^2(m_B+m_2)}{2m_b(m_s+m_q)m_2}\frac{A_1}{A_0}
 +\frac{m_1^2(2m_B^2-(P_B\cdot p_1))}{2m_b(m_s+m_q)m_2(m_B+m_2)}\frac{A_2}{A_0}
\nonumber \\
 & & \hskip0.9cm
 +\frac{3(P_B\cdot p_1)-2m_B^2-m_1^2}{2m_bm_2}\frac{g_+}{A_0}
 +\frac{m_1^2-(P_B\cdot p_1)}{2m_bm_2}\frac{g_-}{A_0}
 +\frac{m_B^2m_1^2-(P_B\cdot p_1)^2}{m_bm_B^2m_2}\frac{h}{A_0} \Biggr) \Biggr],
\label{eq:tilSBV}
\end{eqnarray}
with
%where $g_+$, $g_-$ and $h$ are given by
\begin{eqnarray}
  g_+ &=&
      \frac{1}{2}\Biggl[ \frac{m_B+m_2}{m_B}A_1
     +\frac{m_B^2-m_2^2+m_1^2}{m_B(m_B+m_2)}V \Biggr], \ \ \ \ \
  g_- =
      \frac{1}{2}\Biggl[ \frac{m_B+m_2}{m_B}A_1
     -\frac{3m_B^2+m_2^2-m_1^2}{m_B(m_B+m_2)}V \Biggr],
\nonumber \\
 h &=&
      \frac{m_B\;V}{m_B+m_2}-\frac{m_B\; A_2}{2(m_B+m_2)}
     -\frac{m_Bm_2\;A_0}{m_1^2} +\frac{m_B(m_B+m_2)\; A_1}{2m_1^2}
     -\frac{m_B(m_B-m_2)\; A_2}{2m_1^2},
\end{eqnarray}
where $A_{0,1,2}$ and V are the $B\to V$ form factors \cite{NF,FFLF,FFLCSR}.

\subsection{$B\to VV$ decays}

For longitudinally polarized state with $B\to V_2$ transition, one
has
\begin{eqnarray}
\widetilde S_{V_1V_2}
 &=& -\frac{8}{9}\Biggl[\Biggl(\frac{x(m_B+m_2)}{2}-\frac{m_1m_2}{m_B+m_2}
                               \frac{A_2}{A_1}(x^2-1)\Biggr)
                        \biggl(1 - \frac{m_s}{m_b}\biggr)
 +\frac{1}{2m_b}\frac{g_+}{A_1}\left(m_1m_2(x^2-1)-m_B^2 x\right)
\nonumber \\
 & & \hskip0.9cm
 -\frac{m_B p_c x}{2m_b}\frac{g_-}{A_1}
 -\frac{m_1^2m_2^2}{2m_bm_B^2}\frac{h}{A_1} (x^2-1)\left(\sqrt{x^2-1}+x\right)
%\nonumber \\
% & & \hskip0.9cm
 +\frac{m_1m_2^2}{m_b(m_b+m_q)}\frac{A_0}{A_1}(x^2-1) \Biggr],
 \label{eq:tilSBVV}
\end{eqnarray}
where $x\equiv \frac{p_1\cdot p_2}{m_1m_2}$, and $p_c$ is the c.m.
momentum of the final state particle.
\end{widetext}

%%%%%%%% %%%%%%% %%%%%%% %%%%%%%
%%%%%%%% %%%%%%% %%%%%%% %%%%%%%
%%
%% references
%%


\begin{thebibliography}{99}
\bibitem{note}
This follows Eq.~(\ref{eq:SKphoto}) and is opposite in sign
convention to Ref.~\cite{CHNPRL04}, which follows Ref.~\cite{AGS}.
\end{thebibliography}
\end{document}